# On the nature of filamentary superconductivity in metal-doped hydrocarbon organic materials

D. Hillesheim, K. Gofryk, and A. S. Sefat*

Oak Ridge National Laboratory, Oak Ridge, TN 37831, USA

*corresponding author: *sefata@ornl.gov*

**High temperature superconductivity in K-doped 1,2:8,9-dibenzenopentacene ($C_{30}H_{18}$) has been recently reported [1] with $T_c$ = 33 K, the highest among organic superconductors at ambient pressure. Here we report on our search for superconductivity in K, Ba, and Ca-doped hydrocarbon organic materials. We find that Ba-anthracene ($C_{14}H_{10}$) and K-Picene ($C_{22}H_{14}$) show features characteristics of superconducting state, although very weak. The data suggests that Ba-anthracene might be a new organic superconductor with $T_c$ ~ 35 K.**

Since the discovery of superconductivity, more than a century ago, many groups of different superconducting materials have been discovered and studied. Currently the biggest hope for applications is in high temperature superconducting materials namely cuprates [2], Fe-based pnictides [3] and chalcogenides [4], and organic superconductors [5]. The organic superconductors were first proposed by W. A. Little [6,7] and are light, less dense, lack toxic elements, and are expected to have great potential for the fine-tuning of electrical properties of π-electron networks by metal doping. These materials consist of open-shell molecular units resulting in partial oxidation and reduction of the donor and acceptor molecules. The unpaired electron from the π-molecular orbital (π-hole) of the donor unit is responsible for the electronic properties of these systems. The overlapping of π-orbitals between molecules is causing the π-hole to become itinerant, giving rise to metallic conductivity. Several different types of organic superconductors have been identified and characterized such as the quasi-one-dimensional Bechgaard salts [8], acenes [9] polythiophene [10], fullerens [11], nanotubes [12], and picene [13].

Reports of metal/organic superconductors prepared in liquid ammonia were of interest because of concurrent work involving ammonothermal intercalation of alkali- and alkaline-earth-metals into binary metallic systems. Recently, high-temperature superconductivity (HTS) has been reported in K-doped polycyclic-aromatic-hydrocarbons (PAH) such as picene [13,14], phenanthrene [15], and dibenzopentacene [1]. Furthermore, superconducting state has been also reported in Ba and Sr-doped phenanthrene [16]. Because of the very small shield fractions and non-Meissner feature in most of these organic superconductors, we perform our own investigative work for HTS in such simple PAH. Also, since it was suggested that $T_c$ may increase with longer length of PAH chain [1] and in contrast to theoretical predictions [17], we investigate potassium doping in different-length PAH. Here we present our studies on synthesis and magnetic properties of Ba- and/or Ca-doped anthracene ($C_{14}H_{10}$), phenanthrene ($C_{14}H_{10}$), benzanthracene ($C_{18}H_{12}$), and also K-doped picene ($C_{22}H_{14}$), chrysene ($C_{18}H_{12}$), and 1,2:8,9-dibenzopentacene. We used a hybrid combination of synthesis parameters from the Xue *et. al.* study [1] and the ammonothermal method. Much to our dismay, we did not find clear superconducting signatures in the materials studied, however for Ba-Anthracene, the magnetic susceptibility shows an anomaly at 35 K that may indicate the presence of the very small fraction of a superconducting phase, even though there is no diamagnetic signal. These results together with previously published data could indicate a high sensitivity of the filamentary superconducting state in the organic materials to the preparation details and leave an open question on the origin of the metallic state and bulk superconductivity in these materials.

To prepare Ba and Ca samples we have followed the preparation methods originally proposed for K-based 1,2:8,9-dibenzopentacene [1]. All powder manipulations are performed in air-free or inert gas environments (helium glove box with <1ppm $H_2O$ and <1ppm $O_2$). Typically, a 125 mL Schlenk tube equipped with a stir bar is filled with 100 mg of polycyclic aromatic hydrocarbon and a calculated amount of solid metal, ranging from 0.5 to 3.45 molar equivalents. The loaded tube is evacuated for 1 hour to <50 mtorr. Approximately ~25 mL (at room temperature) of ammonia that has been dried on potassium is vacuum transferred to the reaction vessel with using liquid nitrogen. The mixture is allowed to slowly rise to room temperature and stirred for 6 hours. During this time ~8 bar (6000 torr) pressure are present in reaction flask [18]. The reaction vessel is frozen with liquid nitrogen and the residual solvent is removed via vacuum transfer to produce a dry powder. After each ammonothermal reaction, before the solvent was removed, the mixture was sampled at liquid nitrogen temperatures. No pressure from permanent gas was recorded. The magnetic properties have been studied using a commercial Quantum Design PPMS-7 SQUID magnetometer, with the magnet being reset before each sample insertion. All samples have been screened by the low temperature magnetic susceptibility measurements. Due to high sensitivity of the organic products to air, a special encapsulation procedure of the products in straws was used.

Except for of Ba-anthracene and $K_x$Picene none of the samples showed a clear signature of superconductivity down to 2 K. All the results obtained are summarized in Table 1. Fig.1a shows $\chi(T)$ of Ba-anthracene measured at zero-field-cooled and field-cooled regimes. The curve has characteristic feature of a potentially superconducting material with $T_c \sim 35$ K. The value of the transition temperature is slightly higher than 33 K, recently reported in K-1,2:8,9-dibenzopentacene [1]. However, a diamagnetic signal is not observed in this sample, indicating that the fraction of the superconducting material is very small. From our K-doped materials, only $K_x$Picene shows sign of superconductivity (see Fig.1b). In agreement with published data the superconducting temperature is around 20 K for this material [13,14]. A diamagnetic signal, however, is observed only below ~13 K. Also, similarly to reference [13], the superconducting volume fraction is very small (…%) with no Meissner fraction.

In summary, we have synthesized and studied several alkali-metal- and alkali-earth-doped hydrocarbon samples. Only Ba-anthracene shows weak superconducting features such as irreversibility in zero-field-cooled and field-cooled measurements that may be characteristic of a superconducting state anomaly at 35 K. However, the diamagnetic signal has not been observed indicating that the amount of the superconducting phase in this sample has to be very small. This may suggest that Ba-Anthracene is a new organic superconductor with $T_c \sim 35$ K, hence different amounts of barium doping and also thermal-annealing studies may be performed. Interestingly, all alkali-metal- and alkali-earth-doped hydrocarbon superconductors show a very small superconducting volume fraction [1,13]. In addition, we do not observe any sign of superconductivity (using the same synthesis method as reported in reference [1]) in K-doped 1,2:8,9-dibenzopentacene and Ba-phenanthrene, which are reported to have $T_c \sim 33$ and 5.4 K, respectively [1,15]. This may indicate a high sensitivity of the filamentary superconducting state to the small changes of metal dopant concentration (x). Furthermore, the formation of the metallic superconducting state in these materials is still unclear from our study and also reference [19], requires further studies. In addition, we cannot make any general conclusions about a general trend between the variable lengths of PAH chains to superconducting critical transition temperature.

**Table 1:** The ammonothermal synthesis of K-, Ba-, and Ca-doped PAH. The superconductivity feature is accessed by any feature in magnetic susceptibility down to 2 K. The molar equivalents of each metal to PAH (eq) is indicated.

| Metal | PAH | Superconductivity |
|---|---|---|
| 1 eq Ca | anthracene ($C_{14}H_{10}$) 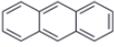 | No |
| 1 eq Ba | anthracene ($C_{14}H_{10}$) 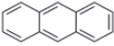 | Anomaly at 35 K |
| 1 eq Ba | phenanthrene ($C_{14}H_{10}$) 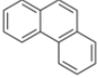 | No |
| 1 eq Ba | Benzanthracene ($C_{18}H_{12}$) 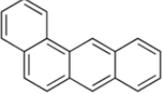 | No |
| 3.45 eq K | chrysene ($C_{18}H_{12}$) 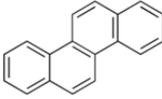 | No |
| 3.45 eq K | picene ($C_{22}H_{14}$) 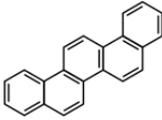 | Anomaly at 22 K |
| 3.45 eq K | 1,2:8,9-dibenzopentacene ($C_{30}H_{18}$) 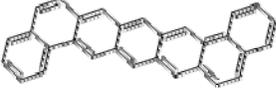 | No |

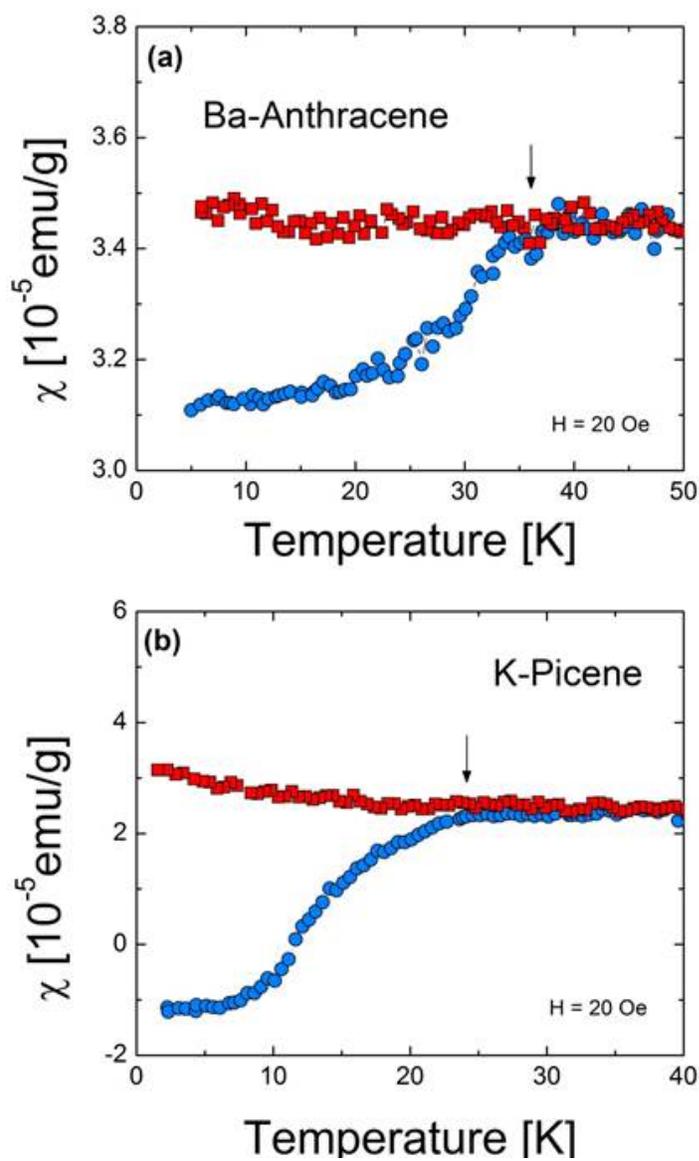

Fig.1. The temperature dependence of the magnetic susceptibility of Ba-anthracene (a) and K-Picene (b). The data were measured in zero-field-cooled (blue circles) and field-cooled (red squares) regimes. Arrows mark anomalies in $\chi(T)$ (see text).